\documentclass[preprintnumbers,nofootinbib,amsmath,amssymb,preprint,aps,prd]{revtex4-1}

\begin{document}
\preprint{IPMU11-0056} \vskip 0.2in
\title{Ho\v{r}ava-Lifshitz gravity with $\lambda\to\infty$}
\author{A. Emir G\"umr\"uk\c{c}\"uo\u{g}lu}
\author{Shinji Mukohyama}
\affiliation{Institute for the Physics and Mathematics of the Universe (IPMU)\\ 
The University of Tokyo, 5-1-5 Kashiwanoha, Kashiwa, Chiba 277-8582, Japan}

\date{\today}

\begin{abstract}
 In the framework of the power-counting renormalizable theory of
 gravitation, recently proposed by Ho\v{r}ava, we study the limit
 $\lambda\to\infty$, which is arguably the most natural candidate
 for the ultraviolet fixed point of the renormalization group flow.
 In the projectable version with dynamical critical exponent $z=3$,
 we analyze the Friedmann-Robertson-Walker background driven by
 the so-called ``dark matter as integration constant'' and
 perturbations around it. We show that amplitudes of quantum
 fluctuations for both scalar and tensor gravitons remain finite
 in the limit and that the theory is weakly coupled under a
 certain condition.
\end{abstract}

\maketitle

\section{Introduction}

A new theory of gravitation proposed recently by
Ho\v{r}ava~\cite{Horava:2009uw} has been attracting significant 
interest. (See \cite{Mukohyama:2010xz,Sotiriou:2010wn} for a review.)
This theory, often called Ho\v{r}ava-Lifshitz (HL) gravity, is
power-counting renormalizable thanks to the anisotropic scaling in the
UV, 
\begin{equation}
 t \to b^z t, \quad \vec{x} \to b \vec{x}, \label{eqn:scaling}
\end{equation}
with the dynamical critical exponent $z\geq 3$.

The scaling (\ref{eqn:scaling}) treats the time and the space in a
different way. Hence, in order to realize this anisotropic scaling, the
$4$-dimensional diffeomorphism invariance cannot be a fundamental 
symmetry of the theory at high energy. Instead, the theory is 
invariant under the so-called foliation-preserving diffeomorphism: 
\begin{equation}
 t \to t'(t), \quad \vec{x} \to \vec{x}\,'(t,\vec{x}). 
\end{equation}
Because of this symmetry, the time kinetic Lagrangian for gravitons is a
linear combination of $K^2$ and $K^{ij}K_{ij}$, where $K_{ij}$ is the
extrinsic curvature of constant-time hypersurfaces and $K=K^i_i$. Thus,
the corresponding terms in the gravitational action are 
\begin{equation}
 I_g \ni \frac{M_{\rm Pl}^2}{2}\int Ndt \sqrt{g}d^3\vec{x}
  \left(K^{ij}K_{ij}-\lambda K^2\right), 
  \label{eqn:time-kinetic-terms}
\end{equation}
where $\lambda$ is a constant. In general relativity (GR) the 
$4$-dimensional diffeomorphism invariance fixes the value of $\lambda$
to $1$. On the other hand, in HL gravity, any value of $\lambda$ is
consistent with the foliation-preserving diffeomorphism invariance.

HL gravity includes not only two degrees of freedom of usual tensor
graviton but also one extra degree of freedom, dubbed the scalar
graviton. The nature of this scalar degree depends on the value of the
parameter $\lambda$. For $1/3<\lambda<1$, the scalar graviton has 
a wrong-sign time kinetic term (i.e. it is a ghost) and thus, this
region is forbidden. For $\lambda<1/3$ or $\lambda>1$, the scalar
graviton has a positive time kinetic term but has a negative sound speed 
squared, 
$c_s^2=-(\lambda-1)/(3\lambda-1)<0$~\cite{Wang:2009yz,Blas:2009qj,Koyama:2009hc}. 
The condition under which the associated long-distance instability does
not show up is~\cite{Mukohyama:2010xz} 
\begin{equation}
 0 < \frac{\lambda-1}{3\lambda-1}
  < \max\left[ \frac{a^2\,H^2}{k^2}, |\Phi|\right]
  \quad \mbox{for}\quad
  H < \frac{k}{a} < \min\left[M_s,\frac{1}{0.01\,{\rm mm}}\right],
  \label{eqn:csbound}
\end{equation}
where $k$ is the comoving momentum scale of interest, $a$ is the scale factor, $H$ is the Hubble
expansion rate of the background cosmology, $M_s$ is the energy scale at
which the anisotropic scaling becomes important for the scalar graviton,
and we have introduced the Newtonian potential $\Phi$ by 
$M_{\rm Pl}^2\, (k/a)^2\,\Phi\sim -\rho$. Here, $\rho$ is the energy density of the
background. This condition essentially says that $\lambda$ must be
sufficiently close to $1$ in the infrared (IR).

The condition (\ref{eqn:csbound}) should be considered as a
phenomenological constraint on properties of the renormalization group
(RG) flow since $\lambda$ is subject to running under the RG flow and in
general, should depend on $k$, $H$ and $\Phi$. This suggests that, in 
order for the theory to be phenomenologically viable, $\lambda=1$ should
be an IR fixed point of the RG flow and that $\lambda$ should approach
$1$ from above sufficiently quickly as the energy scale of the system is 
lowered. In this sense, $\lambda=1+0$ is a candidate for the IR fixed
point of the RG flow. Since the interval $1/3<\lambda<1$ is forbidden, a natural candidate for the UV fixed point that is consistent with the arguments for the IR fixed point above, is $\lambda=+\infty$.

The goal of this paper is to investigate some properties of the
projectable version of the theory without detailed balance, in the
vicinity of the expected UV fixed point, $\lambda=+\infty$. One might
expect a loss of theoretical control in this limit since the coupling
constant diverges. On the contrary, we show below that the  theory is
totally well-behaved and actually simpler in this limit.

The rest of this paper is organized as follows. In
Sec.~\ref{sec:review}, we review the basic equations in HL gravity with
projectability condition. In Sec.~\ref{sec:FRW} we discuss the
background evolution of a FRW geometry describing our local patch of the
universe populated by a perfect fluid. In Sec.~\ref{sec:perturbation} we
discuss the dynamics of tensor and scalar perturbations around the FRW
universe. We conclude with Sec.~\ref{sec:summary} where we 
summarize our results and discuss some of the standing issues. A simple
system of a Lifshitz scalar in HL gravity is investigated in the
Appendix~\ref{app:scalar}.

\section{Ho\v{r}ava-Lifshitz gravity: Review and basic equations}
\label{sec:review}

HL gravity, being a less restricted theory than GR, requires the
temporal and spatial coordinates to be treated on different grounds. The
theory itself is invariant under the so-called foliation-preserving
diffeomorphism, which is a combination of global time reparametrizations
and spatial diffeomorphisms, characterized by the following
infinitesimal transformations 
\begin{equation}
\delta t = f(t)\,,\quad
\delta x^i = \xi^i(t,\vec{x})\,.
\end{equation}
Due to the different scaling dimensions of time and space coordinates,
the $4$-dimensional spacetime metric is not a fundamental
quantity. Instead, fundamental quantities in the HL gravity are the
lapse function $N(t)$, the shift vector $N^i(t,\vec{x})$ and the
$3$-dimensional spatial metric $g_{ij}(t,\vec{x})$. It is still useful,
at least at low energies, to combine them into a $4$-dimensional metric
in the fashion of ADM \cite{Arnowitt:1962hi},
\begin{equation}
ds^2= -N^2\,dt^2 +g_{ij}\,
 \left(dx^i+N^i dt\right)\left(dx^j+N^j dt\right)\,.
\end{equation}
Note that the shift vector $N^i$ and spatial metric $g_{ij}$ depend on
all four coordinates but that the lapse function $N$ is assumed to be a
function of time only. The latter assumption, dubbed {\it the
projectability condition} is consistent with the foliation preserving
diffeomorphism in the sense that a projectable $N$ is mapped to another
projectable $N$.

Starting with the time kinetic action (\ref{eqn:time-kinetic-terms}),
the most general gravitational action that respects the symmetries of
the theory can be constructed as 
\begin{equation}
I_{g} = \frac{M_{\rm Pl}^2}{2} \int N\,dt \, \sqrt{g}\, d^3 \vec{x}
  \left(K^{ij}K_{ij} - \lambda K^2 -2\Lambda + R
   +L_{z>1}\right),
\label{initact}
\end{equation}
where 
\begin{equation}
K_{ij} \equiv \frac{1}{2\,N} 
 \left(\dot{g}_{ij}-D_i N_j - D_j N_i \right) 
\end{equation}
is the extrinsic curvature and its trace $K$ is obtained with its
contraction with the $3$d induced metric. The broken Lorentz symmetry
manifests itself as an arbitrary parameter $\lambda$, which acquires the
value $1$ in GR. In the above action, we fixed the coefficient of the
scalar curvature to unity by a choice of unit so that the
Einstein-Hilbert action is reproduced in the IR limit with
$\lambda\rightarrow 1$. Finally, the part of the action denoted by
$L_{z>1}$ contains the higher spatial derivative terms and controls the
UV behavior of the system. For definiteness, we will focus on the case
with $z=3$ scaling in the UV in the remainder of this paper. If the
detailed balance is not enforced but if the spatial parity and time
reflection symmetries are imposed, this choice allows spatial derivative
terms up to sixth order as
\begin{eqnarray}
 \frac{M_{Pl}^2}{2}L_{z>1} &=& 
  \left(c_1\,D_i R_{jk} D^i R^{jk} 
   + c_2\, D_i R\,D^i R +c_3 R^j_i R^k_j R^i_k 
   + c_4 R\,R^j_i R^i_j +c_5 R^3 \right) \nonumber\\
 && +\left(c_6 R^j_i R^i_j +c_7 R^2\right)\,,
\end{eqnarray}
where $D_i$ is the covariant derivative with respect to the $3$d metric
and $c_i$ are constants.

The effect of matter on the dynamics is provided by the additional
action term $I_m$, which is also required to be invariant under the 
foliation-preserving diffeomorphism.

By variation with respect to $g_{ij}(t,x)$, we obtain the dynamical
equation 
\begin{equation}
 {\cal E}_{g ij}+{\cal E}_{m ij}=0, \label{vargij}
\end{equation}
where
\begin{equation}
 {\cal E}_{g ij} \equiv g_{ik}g_{jl}\frac{2}{N\sqrt{g}}
  \frac{\delta I_g}{\delta g_{kl}}, \quad
 {\cal E}_{m ij} \equiv g_{ik}g_{jl}\frac{2}{N\sqrt{g}}
  \frac{\delta I_m}{\delta g_{kl}}
  = T_{ij}. 
\end{equation}
Note that the matter sector (as well as the gravity sector) should
be invariant under spatial diffeomorphism (as a part of the foliation 
preserving diffeomorphism) and thus it makes sense to define $T_{ij}$ in
general. The explicit expression for ${\cal E}_{g ij}$ is given by 
\begin{eqnarray}
 {\cal E}_{g ij} & = & M_{Pl}^2
  \left[
  -\frac{1}{N}(\partial_t-N^kD_k)p_{ij}
  + \frac{1}{N}(p_{ik}D_jN^k+p_{jk}D_iN^k)
  \right.\nonumber\\
 & & \left.
  - Kp_{ij} + 2K_i^kp_{kj}
      + \frac{1}{2}g_{ij}K^{kl}p_{kl}+ \frac{1}{2}\Lambda g_{ij} 
      - G_{ij}\right]
 + {\cal E}_{z>1 ij}, \nonumber\\
  p_{ij} & \equiv & K_{ij} - \lambda Kg_{ij},
\end{eqnarray} 
where ${\cal E}_{z>1 ij}$ is the contribution from $L_{z>1}$ and
$G_{ij}$ is Einstein tensor of $g_{ij}$.

Variation with respect to the shift $N^i(t,x)$ leads to the momentum
constraint 
\begin{equation}
 {\cal H}_{g i}+{\cal H}_{m i}=0, 
  \label{varni}
\end{equation}
where
\begin{equation}
 {\cal H}_{g i} \equiv - \frac{\delta I_g}{\delta N^i}
  = -M_{Pl}^2\sqrt{g}D^jp_{ij}, \quad
 {\cal H}_{m i} \equiv 
 -\frac{\delta I_m}{\delta N^i}. 
\end{equation}
The only remaining equation is the Hamiltonian constraint, obtained by
variation with respect to the lapse function $N(t)$, 
\begin{equation}
 H_{g\perp}+H_{m\perp}=0, \label{hamcongen}
\end{equation}
where
\begin{equation}
 H_{g\perp} \equiv -\frac{\delta I_g}{\delta N}
  = \int d^3\vec{x} {\cal H}_{g\perp}, \
 H_{m\perp} \equiv -\frac{\delta I_m}{\delta N},
\end{equation} 
and
\begin{equation}
  {\cal H}_{g\perp} = \frac{M_{Pl}^2}{2}\sqrt{g}
  (K^{ij}p_{ij}-\Lambda-R-L_{z>1}). 
\end{equation}
Here, we stress that due to the projectability condition, which
restricts $N$ to be only time dependent, the Hamiltonian constraint in
HL gravity is a global one, in contrast to the local one in GR.

Just for comparison, in a Lorentz invariant theory, the energy-momentum
tensor is defined as 
\begin{equation}
T_{\mu\nu}^{\rm (LI)} = 
 -\frac{2}{\sqrt{-^{(4)}g}}\,
 \frac{\delta I_m^{\rm (LI)}}{\delta g^{\mu\nu}}\,,
\label{enmomdef}
\end{equation}
and thus the matter terms in the constraints are expressed as
\begin{equation}
 H_{m\perp} = \int d^3\vec{x}\sqrt{g}\ 
  T^{(LI)}_{\mu\nu}n^{\mu}n^{\nu}, 
  \quad 
 {\cal H}_{m i} = \frac{1}{\sqrt{g}}T^{(LI)}_{i\mu}n^{\mu},
\end{equation}
where we defined the $4$-vector $n^\mu$ to be the unit vector normal to
the constant-time hypersurfaces, with 
\begin{equation}
n^\mu \partial_\mu \equiv 
 \frac{1}{N}\,\left(\partial_t - N^i\,\partial_i\right)\,.
\end{equation}

\section{FRW background}
\label{sec:FRW}

A Friedmann-Robertson-Walker (FRW) metric, 
\begin{equation}
ds^2 = -dt^2 + a(t)^2 \,d{\vec x}^2\, ,
\end{equation}
is supposed to describe the large-scale, overall behavior of the
geometry in our local patch of the universe. Since the universe far 
outside the present horizon may be very different from the local
universe inside the horizon, we should not expect the same FRW geometry 
to describe the whole spacetime including the region far outside our
local patch~\footnote{This is in accord with the so called gradient
expansion approach to super-horizon nonlinear cosmological
perturbations~\cite{Mukohyama:soon}.}. Nonetheless, in general
relativity, since the Hamiltonian constraint is a local equation
satisfied at each spatial point, it leads to a Friedmann equation 
applicable to our local patch of the universe. On the other hand, in 
HL gravity with the projectability condition, the Hamiltonian constraint
is a global equation integrated over the whole space. For this reason,
the Hamiltonian constraint in HL gravity does not tell anything useful
about the ``local'' FRW geometry~\cite{Mukohyama:2009mz}.

Therefore, in HL gravity with the projectability condition, we do not
have a Friedmann equation applicable to our local FRW universe. Instead,
we have the dynamical equation (\ref{vargij}) in the form
\begin{equation}
-\frac{3\,\lambda - 1}{2}\,\left(2\,\dot{H} + 3\,H^2\right)
 = \frac{P}{M_{\rm Pl}^2}\,,
\label{bgeq}
\end{equation}
where, in accord with the ``local'' homogeneity and isotropy of the
``local'' FRW geometry, we have assumed that the stress tensor of matter
is ``locally'' homogeneous and isotropic as
\begin{equation}
 T_{ij}= P(t)g_{ij}\quad {\cal H}_{m i} =0\,. \label{enmom1}
 \end{equation}
For the same reason as why we do not have a Friedmann equation
applicable to a ``local'' FRW universe, we do not have a conservation
equation for the ``locally'' homogeneous and isotropic
matter. Therefore, we define a quantity $Q(t)$ by~\footnote{
In HL gravity  the projectability condition implies that we do not have
to define a ``local'' energy density $\rho$. Nonetheless, just for our
convenience we can still define $\rho$ by pretending as if $N$ were a
function of time and spatial coordinates. With $\rho$ defined in this
way, the quantity $Q$ measures the amount of deviation from what we
would expect in theories with $4$-dimensional spacetime
diffeomorphism.}
\begin{equation}
\dot{\rho}+3\,H\,\left(\rho+P\right)=- Q\,,
\label{encon}
\end{equation}
where $H \equiv \dot{a}/a$. Note that $Q$ is generically nonzero at high
energies. From Eqs.(\ref{bgeq}) and (\ref{encon}), one can find a
generalized Friedmann equation \cite{Mukohyama:2010xz} 
\begin{equation}
 3\,M_{\rm Pl}^2\,H^2 = \rho_{''{\rm dm}''}
  + \frac{2\,}{3\,\lambda-1}\,\rho\,,
\label{genfr}
\end{equation}
where
\begin{equation}
\rho_{''{\rm dm}''} \equiv \frac{1}{a^3} 
 \left[ C_0 + \frac{2}{3\,\lambda-1} \int_{t_0}^t Q(t')\,a^3(t')dt'\right]\,,
\label{cdef}
\end{equation}
with $C_0$ an integration constant. The quantity $\rho_{''{\rm dm}''}$ 
is the ``dark matter as integration constant'', associated with the lack
of a local Hamiltonian constraint in HL gravity. See
ref.~\cite{Mukohyama:2009mz} for more general cases.

Equation of motion of matter leads to conservation equation at
least at low energy, provided that the local Lorentz invariance is
restored in the matter sector as required by many experimental and
observational data. In this case,  we have $Q\to 0$ as $a\to\infty$, and
the integral part of Eq.(\ref{cdef}) converges to a constant. Thus at
low energy, the $\rho_{\rm ''dm''}$ component redshifts like
nonrelativistic matter, or pressureless dust.

Let us now consider the limit $\lambda\to +\infty$. The dynamical
equation (\ref{bgeq}) is reduced to 
\begin{equation}
 2\,\dot{H} + 3\,H^2 = 0\,,
  \label{bgeq-simple}
\end{equation}
and the generalized Friedmann equation (\ref{genfr}) is greatly
simplified as 
\begin{equation}
 3\,M_{\rm Pl}^2\,H^2 = \rho_{''{\rm dm}''}\,, \quad
  \rho_{''{\rm dm}''} \equiv \frac{C_0}{a^3}\,.
\label{genfr-simple}
\end{equation}
This shows that the matter sector decouples from the gravity sector and
that the evolution of the local FRW universe is dominated by the ``dark
matter as integration constant'' in the limit $\lambda\to\infty$.

However, from the cosmological viewpoint, we need to specify what we
exactly mean by the limit $\lambda\to\infty$. Supposing that
$\lambda=+\infty$ is a UV fixed point of the RG flow, the second term in  
the r.h.s. of (\ref{genfr}) is indeed suppressed by $1/\lambda$ in the
early universe. A similar suppression of coupling to the matter sector
can be observed for the integral term in (\ref{cdef}). However, the
increase in $\lambda$ going earlier in time does not necessarily imply
that the matter sector is decoupled from gravity since  $\rho$ (and $Q$)
also becomes large in the early universe. In order to obtain the
decoupled equation (\ref{genfr-simple}), what we really have to ensure
is that 
\begin{equation}
 \frac{\lambda\,\rho_{''{\rm dm}''}}{\rho}
  \simeq  \frac{\lambda\,C_0}{\rho\, a^3}
  \gg 1. \label{eqn:condition-large-lambda}
\end{equation}
Assuming logarithmic running of the coupling $\lambda \sim \log (H/M)$
for $H\gg M$, if the fluid energy redshifts faster than $a^{-3}$, the
fluid generically dominates the expansion in the asymptotic past. On the
other hand, even if the fluid dominates the expansion early on, the
``dark matter'' energy can in principle catch up later and become the
dominant source while the theory is still in the UV regime. In this
case, even though the earlier evolution exhibits a coupled behavior, the
modes that are deep inside the horizon at the time of transition will
not carry any memory of this early behavior. We will focus on scales for
which at the time of (sound) horizon crossing the UV behavior 
$\lambda \gg 1$ is still valid and the fluid contribution to the
expansion is suppressed relative to the ``dark matter'' as in
(\ref{eqn:condition-large-lambda}). Any lengths beyond this scale are
assumed to be well beyond the current observable universe.

As an alternative case, we can also consider a situation in which the
fluid is pressure-less. In this case, the ratio 
$\lambda\,\rho_{''{\rm dm}''}/\rho$ grows logarithmically in the UV 
direction, and the matter indeed decouples from geometry in the
asymptotic past~\footnote{
For time scales in which the logarithmic running of $\lambda$ is not
appreciable, there is no distinction between this pressure-less fluid and
the ``dark matter as integration constant'' at the background level,
since both have the same equation of state. On the other hand, at the
level of perturbations, they are distinct even without taking into
account the running of $\lambda$ since the rest frame
of the ``dark matter'' (but not that of the fluid) is synchronized with
the spacetime foliation and dispersion relations for all physical
degrees of freedom are associated with this
foliation~\cite{Mukohyama:2009tp}.}.

\section{Perturbation}
\label{sec:perturbation}

In this section, motivated by the decoupling between gravity and matter
in the limit $\lambda\to\infty$ observed in the previous section, we
study a pure gravity system and analyze the evolution of perturbations
around the FRW background driven by the ``dark matter as integration
constant''. (In Appendix~\ref{app:scalar}, in order to justify this
treatment we consider a scalar field in HL gravity and show that gravity
and matter are decoupled in the limit $\lambda\to\infty$ for linear
perturbations.) We investigate the UV regime with the dynamical critical
exponent $z=3$ and show that the amplitude of quantum fluctuations
remains finite and that the system is well behaved in the
$\lambda\to\infty$ limit.

One of the most important properties of HL gravity is the anisotropic
scaling (\ref{eqn:scaling}) with $z\geq 3$ since the power-counting
renormalizability stems from it. Intriguingly, with the minimal value
$z=3$, this scaling can lead to a mechanism to generate scale-invariant
cosmological perturbations even without
inflation~\cite{Mukohyama:2009gg}. Let us briefly review this mechanism
before going into the detailed analysis of perturbations.

With $z=3$, we would like to know the condition for generation of
super-horizon cosmological perturbations. Generation of super-horizon
cosmological perturbation is nothing but oscillation followed by
freeze-out. Each mode oscillates for $\omega^2\gg H^2$ and freezes out
for $\omega^2\ll H^2$, where $\omega$ is the frequency of a mode of
interest and $H=\dot{a}/a$ is the Hubble expansion rate. Thus, the
condition for generation of cosmological perturbations is 
$\partial_t(H^2/\omega^2)>0$. With the dispersion  
relation $\omega^2\simeq (\vec{k}^2/a^2)^3/M^4$ expected from the 
$z=3$ scaling, where $\vec{k}$ is the comoving momentum and $M$
is a characteristic mass scale, this condition is reduced to
$\partial_t^2(a^3)>0$ for an expanding universe. This condition can be
satisfied by, for example, a power-law expansion $a\propto t^p$ with
$p>1/3$, and does not require accelerated expansion ($p>1$),
i.e. inflation.

For concreteness, let us consider a scalar field $\phi$ with a canonical
time kinetic term. The anisotropic scaling (\ref{eqn:scaling}) implies
that $\phi$ should scale as 
\begin{equation}
 \phi \to b^{-s} \phi, \quad s = \frac{3-z}{2}.
\end{equation}
From this, it is expected that the amplitude of quantum fluctuations in
a FRW background should be
\begin{equation}
 \delta\phi \sim M \times \left(\frac{H}{M}\right)^{\frac{3-z}{2z}},
\end{equation}
where $M$ is a characteristic mass scale in the action for $\phi$,
e.g. the scale suppressing higher spatial derivative terms. This 
reproduces the well-known result $\delta\phi\sim H$ for Lorentz 
invariant theories ($z=1$) and $\delta\phi\sim (M^3H)^{1/4}$ for ghost 
inflation~\cite{ArkaniHamed:2003uz} ($z=2$). For HL gravity with $z=3$,
we have a Hubble-independent result, $\delta\phi\sim M$. Thus the
amplitude of quantum fluctuations is expected to be scale-invariant in HL
gravity with $z=3$. This also applies to both tensor graviton and scalar
graviton.

While $\delta\phi\sim M$ is generically expected for the HL gravity with
$z=3$, a numerical coefficient in front of $M$ in the right hand side
may depend on $\lambda$. It is not a priori clear whether this numerical
coefficient remains finite or diverges when the $\lambda\to\infty$ limit
is taken. In the following, we shall explicitly show that the amplitudes
for tensor and scalar gravitons indeed remain finite in this limit.

We shall also investigate nonlinear interactions among tensor and scalar
gravitons and show that the system remains weakly coupled in the UV with
$\lambda\to\infty$, provided that 
\begin{equation}
 -c_1 \gg M_{\rm Pl}^{-2}, \quad 
  -(3c_1+8c_2) \gg M_{\rm Pl}^{-2}. 
  \label{eqn:condition-RGflow}
\end{equation}
Since $c_1$, $c_2$ and $M_{\rm Pl}^2$ are subject to running under 
the RG flow, this should be considered as a nontrivial condition on
properties of the RG flow in the UV.

\subsection{Tensor Modes}

We now consider a pure gravity system and analyze tensor perturbations
around the FRW background driven by the ``dark matter as integration
constant''. Let us consider metric perturbations of the form 
\begin{equation}
 \delta N = 0, \quad \delta N_i = 0, \quad
\delta g_{ij} = a^2 h_{ij}\,,
\end{equation}
where $h_{ij}$ is a transverse and traceless $3$d tensor. The part of
the gravitational action (\ref{vargij}) containing the terms quadratic
in tensor degrees can be obtained as 
\begin{equation}
I_g \ni  \frac{M_{\rm Pl}^2}{8} 
 \int dt \, d^3\vec{x} \,a^3 \delta^{ik}\delta^{jl}
 \left[ \dot{h}_{ij} \dot{h}_{kl} + h_{ij}\, 
  {\cal O }_t \, h_{kl} \right]\,.
\label{actiontensor}
\end{equation}
In the above action, the spatial derivatives are contained in the
operator 
\begin{equation}
{\cal O}_t \equiv 
 \frac{1}{a^2} \,\triangle - \frac{\kappa_t}{a^4\,M_t^2} \,\triangle^2 
 + \frac{1}{a^6\,M_t^4} \,\triangle^3\,,
\label{defop}
\end{equation}
where $\triangle \equiv \delta^{ij} \partial_i \partial_j $ and $M_t$ is
some characteristic energy scale defined through 
\begin{equation}
\frac{1}{M_t^4} \equiv -2\frac{c_1}{M_{\rm Pl}^2} \,,\quad 
 \frac{\kappa_t}{M_t^2} \equiv -2\frac{c_6}{M_{\rm Pl}^2}\,.
\end{equation}

We remind the reader that in Eq.(\ref{defop}), the coefficient for the
linear $\triangle/a^2$ term has already been fixed by a choice of unit
(see discussion after (\ref{initact})). Furthermore, we constrain the
sign of the $\triangle^3$ term so that the evolution of the mode is
stable in the UV at the asymptotic past and a vacuum state can be
unambiguously defined. Finally, we do not restrict the sign of
$\kappa_t$, but assume it is of order $1$ in the following just for
simplicity.

We now proceed with the quantization of the tensor mode by first
expanding the tensor degrees in Fourier space as 
\begin{equation}
h_{ij} (t, {\vec x}) = \frac{1}{\left(2\,\pi\right)^{3/2}}\,
 \sum_{\sigma=1,2} \int d^3k \, {\rm e}^{i\,{\vec k}\cdot {\vec x}}  
 \,\epsilon_{ij}^\sigma (\vec{k}) \, \hat{h}_\sigma(t,{\vec k})\,,
\end{equation}
where $\epsilon_{ij}^\sigma$ are the transverse-traceless polarization
tensors and $\sigma$ can take values $1$ or $2$. It is convenient to
introduce a new time parameterization as 
\begin{equation}
dy \equiv \omega_t\,dt\,,
\end{equation}
where $\omega_t$ is the frequency of the form
\begin{equation}
\omega_t^2 \equiv \frac{k^2}{a^2} 
 +\frac{\kappa_t\,k^4}{a^4\,M_t^2} +\frac{k^6}{a^6\,M_t^4}\,.
\label{omegat}
\end{equation}
This brings the kinetic part of the action (\ref{actiontensor}) to
\begin{equation}
I_g \ni \frac{M_{\rm Pl}^2}{8}\sum_\sigma \int dy\,d^3k \,a^3\,\omega_t\,\hat{h}^{\dagger\,\prime}_\sigma \,\hat{h}'_\sigma\,,
\end{equation}
where prime denotes differentiation with respect to the new time $y$. Expanding the operator $\hat{h}$ in a creation/annihilation operator basis as
\begin{equation}
\hat{h}_\sigma(\vec{k}) \equiv h_\sigma (k) \,\hat{a}_\sigma ( \vec{k}) + h^*_\sigma (k) \,\hat{a}^\dagger_\sigma ( -\vec{k})\,,
\end{equation}
we introduce the mode functions that give rise to a canonically normalized kinetic action 
\begin{equation}
\bar{h}_\sigma \equiv \frac{M_{\rm Pl}\,a^{3/2}\,\sqrt{\omega_t}}{2}\,h_\sigma\,,
\end{equation}
obeying the equation of motion
\begin{equation}
\bar{h}''_\sigma + 
 \left[1+ 
  \frac{1}{\omega_t^2}\,
  \left(\frac{3\,\dot{\omega_t}}{4\,\omega_t^2}
   -\frac{\ddot{\omega_t}}{2\,\omega_t}\right)
 \right]\bar{h}_\sigma = 0\,.
\label{eqcanonh}
\end{equation}
Noting that the frequency decreases as $a^{-3}$ in the UV and the pure gravity background satisfies
(\ref{bgeq-simple}), the above equation becomes simply 
\begin{equation}
\bar{h}''_\sigma + \bar{h}_\sigma= 0\,.
\label{eqcanonh2}
\end{equation}
Fixing their amplitude from the kinetic part of the action and requiring
that the corresponding state should minimize the quadratic Hamiltonian
of the system~\cite{Mukohyama:2009gg}, the mode functions of the tensor
field can be written as 
\begin{equation}
h_\sigma = \frac{\sqrt{2}}{\sqrt{\omega_t}\,M_{\rm Pl}\,a^{3/2}} \,{\rm e}^{-i\,y} \simeq\frac{\sqrt{2} M_t}{k^{3/2} \,M_{\rm Pl}} \,{\rm e}^{-i\,y} \,.
\label{hsol}
\end{equation}
Through the two-point function, we define the tensor power spectrum $P_t$ as
\begin{equation}
\sum_\sigma \left\langle \hat{h}_\sigma (t, \vec{k})  \hat{h}^\dagger_\sigma (t, \vec{k}') \right\rangle \equiv \frac{2\,\pi^2}{k^3}\,\delta^{(3)}(\vec{k}-\vec{k}') P_t\,.
\end{equation}
The power spectrum of the modes both sub and super horizon in the UV epoch turns out to be both scale invariant and time independent
\begin{equation}
P_t = \frac{k^3}{2\,\pi^2} \sum_\sigma \left\vert h_\sigma \right\vert^2 = \frac{2}{\pi^2} \left(\frac{M_t}{M_{\rm Pl}}\right)^2\,.
\label{spect}
\end{equation}

\subsection{Scalar Modes}
\label{sec:pfsca}

In this subsection, we discuss the evolution of scalar
perturbations. The metric tensor has three local (i.e. space-dependent)
and one global (i.e. space-independent) scalar degrees of freedom 
\begin{equation}
\delta N = A\,,\quad
\delta N_i = \partial_i B\,,\quad
\delta g_{ij} = a^2\left[2\,\delta_{ij}\,\zeta +\partial_i\partial_j h_L\right]\,,
\end{equation}
where $A=A(t)$ depends only on time in accordance with the
projectability condition discussed in Sec.\ref{sec:review}. We fix the
two scalar gauge degrees of freedom by setting $A=h_L=0$. In 
this convenient gauge, the momentum constraint (\ref{varni}) reads 
\footnote{Although up to this point, we did not make any assumption on
the details of the background evolution or the value of the constant
$\lambda$, the linear equations in this section do not cover the case
$\lambda=1$ due to infinities arising in some of the relations,
e.g. Eq.(\ref{eqB}). However, the existence of such issues does not
necessarily imply that the theory is not continuously connected to
$\lambda=1$ limit, but it is merely a manifestation that the
perturbative expansion breaks down. The concrete study of the continuity
requires a nonlinear analysis and is beyond the scope of the present
paper. See the discussion in Sec.\ref{sec:summary} for further comments
on this issue.} 
\begin{equation}
 B = 
  \frac{3\,\lambda - 1}{\lambda -1}\,
  \frac{\dot{\zeta}}{a^{-2}\triangle}\,,
  \label{eqB}
\end{equation}
while the equations of motion (\ref{vargij}) leads to 
\begin{equation}
\ddot{\zeta}+3\,H\,\dot{\zeta} -{\cal O}_s \zeta = 0\,,
\label{eqz}
\end{equation}
where we defined the operator
\begin{equation}
{\cal O}_s \equiv \frac{\lambda-1}{3\,\lambda-1}\, \left(-\frac{1}{a^2} \,\triangle - \frac{\kappa_s}{a^4\,M_s^2} \,\triangle^2 + \frac{1}{a^6\,M_s^4} \,\triangle^3\right)\,,
\end{equation}
with
\begin{equation}
\frac{1}{M_s^4} \equiv -2\frac{3\,c_1+8\,c_2}{M_{\rm Pl}^2} \,,\quad \frac{\kappa_s}{M_s^2} \equiv -2\frac{3\,c_6+8\,c_7}{M_{\rm Pl}^2}\,.
\end{equation}
Proceeding as in previous subsection, we expand the scalar degrees in
Fourier space, for which Eq.(\ref{eqz}) becomes
\begin{equation}
\ddot{\hat{\zeta}}+3\,H\,\dot{\hat{\zeta}}
 +\omega^2_s \hat{\zeta} = 0\,,
\label{eqz2}
\end{equation}
and the frequency of the scalar graviton perturbation is defined as
\begin{equation}
\omega_s^2 \equiv \frac{\lambda-1}{3\,\lambda-1}\,\left(-\frac{k^2}{a^2} + \frac{\kappa_s\,k^4}{a^4\,M_s^2}  + \frac{k^6}{a^6\,M_s^4} \right)\,.
\end{equation}
The form of the scalar mode frequency implies that at early times,
$\omega_s^2$ is dominated by the (positive) term proportional to $k^6$
and modes are in an oscillatory regime, much like the tensor modes
discussed in the previous subsection. On the other hand, the frequency
at late times may become dominated by the (negative) $k^2$ term,
creating a ground for a linear instability. However, this happens after
the Hubble friction takes over, so the time scale of this instability is
not short enough to have an effect on the evolution. See Eq.(\ref{eqn:csbound}) for the more general condition under which the
long-distance instability does not show up.

We now proceed with the quantization of the scalar graviton
degree. Under time parameterization $dy \equiv \omega_s dt$, the kinetic
part of the scalar action reduces to 
\begin{equation}
I_g \ni M_{\rm Pl}^2 \,\left(\frac{3\,\lambda-1}{\lambda-1}\right)\int dy\,d^3 k\,a^3\,\omega_s \,\hat{\zeta}^{\dagger\,\prime}\,\hat{\zeta}'\,.
\end{equation}
The mode function for the canonical field can then be defined through
\begin{equation}
\bar{\zeta} \equiv \sqrt{2\,\omega_s}\,\sqrt{\frac{3\,\lambda-1}{\lambda-1}}\,a^{3/2}\,M_{\rm Pl}\,\zeta\,,
\end{equation}
with equation of motion
\begin{equation}
\bar{\zeta}'' + 
 \left[1+ \frac{1}{\omega_s^2}\,
  \left(\frac{3\,\dot{\omega_s}}{4\,\omega_s^2}
   -\frac{\ddot{\omega_s}}{2\,\omega_s}\right)
 \right]\bar{\zeta} = 0\,.
\label{eqz3}
\end{equation}
The time dependence of the frequency $\omega_s^2$ is qualitatively the
same as that of the tensor modes. In the UV regime, where one can
approximate $\dot{\omega}_s \simeq -3\,H\,\omega_s$, one obtains a
simple equation for the mode functions 
\begin{equation}
\bar{\zeta}''+\bar{\zeta} = 0\,.
\end{equation}
The canonical scalar modes evolve qualitatively the same as the tensor
modes in (\ref{eqcanonh2}). By going from the canonical mode to the
physical one, the solution for the scalar mode function can be written as 
\begin{equation}
\zeta = \frac{1}{2\,M_{\rm Pl}\,a^{3/2}\,\sqrt{\omega_s}}\,\sqrt{\frac{\lambda-1}{3\,\lambda-1}}\,{\rm e}^{-i y} \simeq
\frac{1}{2\times3^{1/4}\,k^{3/2}} \frac{M_s}{M_{\rm Pl}} {\rm e}^{-i y}\,,
\end{equation}
resulting in a scale invariant scalar spectrum
\begin{equation}
P_s =\frac{k^3}{2\,\pi^2} \left\vert \zeta \right\vert^2 = \frac{1}{4\,\sqrt{3}\,\pi^2} \left(\frac{M_s}{M_{\rm Pl}}\right)^2\,.
\label{specs}
\end{equation}

The tensor-to-scalar ratio for the primordial perturbations thus depends
on the ratio of the two energy scales $M_t$ and $M_s$ through, 
\begin{equation}
\frac{P_t}{P_s} = 8\,\sqrt{3}\,\left(\frac{M_t}{M_s}\right)^2\,.
\end{equation}

\subsection{Cubic Terms}

In this subsection we consider nonlinear perturbations around the FRW
background driven by the ``dark matter as integration constant''. We
adopt the following metric ansatz. 
\begin{equation}
 N = 1, \quad N_i = \partial_i B + n_i, \quad
  g_{ij} = a^2e^{2\zeta}\left(e^h\right)_{ij},
\end{equation}
where $n_i$ is transverse and $h_{ij}$ is transverse traceless:
$\partial^in_i=0$, $\partial^ih_{ij}=0$ and $h^i_{\ i}=0$. Throughout
this subsection, indices are raised and lowered by $\delta^{ij}$ and
$\delta_{ij}$. We consider $\zeta$, $B$, $n_i$ and $h_{ij}$ as
$O(\epsilon)$ and perform perturbative expansion with respect to
$\epsilon$. 

In order to calculate the action up to cubic order, it suffices to solve
the momentum constraint up to the first order. The momentum constraint
at the first order is
\begin{equation}
 \partial_i\left[(3\lambda-1)\,a^2\,\dot{\zeta}-(\lambda-1)\partial^2B\right]
  + \frac{1}{2}\partial^2n_i = 0, 
\end{equation}
leading to 
\begin{equation}
 B = \frac{3\lambda-1}{\lambda-1}\frac{\dot{\zeta}}{a^{-2}\partial^2}, 
  \quad n_i = 0, \label{eqn:constraint-solution}
\end{equation}
where $\partial^2=\partial^i\partial_i$. 

It is somewhat cumbersome but straightforward to calculate the kinetic
action up to the third order. The result is 
\begin{eqnarray}
 I_{kin} & = & \frac{M_{Pl}^2}{2}\int Ndt\sqrt{g}d^3\vec{x}
  (K^{ij}K_{ij}-\lambda K^2) \nonumber\\
 & = & M_{Pl}^2\int dtd^3\vec{x} a^3
  \left[ -\frac{3}{2}(3\lambda-1)H^2
   + \frac{3}{2}(3\lambda-1)(2\dot{H}+3H^2)\zeta
   \left(1+\frac{3}{2}\zeta+\frac{3}{2}\zeta^2\right)
   \right.
   \nonumber\\
 & & 
  \qquad
+ (1+3\zeta)\left(a^{-2}\dot{\zeta}\partial^2B
		+\frac{1}{8}\dot{h}^{ij}\dot{h}_{ij}\right)
   + \frac{1}{2}a^{-4}\zeta\partial^i
   (\partial_iB\partial^2B+3\partial^jB\partial_i\partial_jB)
   \nonumber\\
 & & 
  \left.  \qquad
   + \frac{1}{2}(a^{-2}\partial^kh^{ij}\partial_kB-3\dot{h}^{ij}\zeta)
   a^{-2}\partial_i\partial_jB
   -\frac{1}{4}a^{-2}\dot{h}^{ij}\partial_kh_{ij}\partial^kB
  \right]  + O(\epsilon^4).
\end{eqnarray}
The first term does not depend on the perturbation and the second term,
which is proportional to $2\dot{H}+3H^2$, vanishes because of the
background equation of motion (\ref{bgeq-simple}). Thus what we are
interested in are the quadratic part $I_{kin}^{(2)}$ and the cubic part
$I_{kin}^{(3)}$, where
\begin{eqnarray}
 I_{kin}^{(2)} & = & M_{Pl}^2\int dtd^3\vec{x} a^3
  \left( a^{-2}\dot{\zeta}\partial^2B
		+\frac{1}{8}\dot{h}^{ij}\dot{h}_{ij}\right), 
  \nonumber\\
 I_{kin}^{(3)} & = & M_{Pl}^2\int dtd^3\vec{x} a^3
  \left[
   3\zeta\left(a^{-2}\dot{\zeta}\partial^2B
		+\frac{1}{8}\dot{h}^{ij}\dot{h}_{ij}\right)
   + \frac{1}{2}a^{-4}\zeta\partial^i
   (\partial_iB\partial^2B+3\partial^jB\partial_i\partial_jB)
   \right.
   \nonumber\\
 & & 
  \left.
  \qquad
   + \frac{1}{2}(a^{-2}\partial^kh^{ij}\partial_kB-3\dot{h}^{ij}\zeta)
   a^{-2}\partial_i\partial_jB
   -\frac{1}{4}a^{-2}\dot{h}^{ij}\partial_kh_{ij}\partial^kB
  \right].
\end{eqnarray}
When $B$ is eliminated by using (\ref{eqn:constraint-solution}), one can
easily see that each term in $I_{kin}^{(3)}$ is marginal, i.e. has
vanishing scaling dimension under the scaling (\ref{eqn:scaling}), and 
that each coefficient remains of $O(1)$ (multiplied by the overall
factor $M_{Pl}^2$) in the limit $\lambda\to\infty$. 

As we have already calculated power spectra in the previous subsections,
we know that the amplitudes of quantum fluctuations are
\begin{equation}
 \langle h_{ij}\, h_{kl} \rangle \sim \left(\frac{M_t}{M_{Pl}}\right)^2, \quad
 \langle \zeta\, \zeta \rangle \sim \left(\frac{M_s}{M_{Pl}}\right)^2. 
\end{equation}
Thus, $I_{kin}^{(3)}$ is smaller than $I_{kin}^{(2)}$ and the
perturbative expansion makes perfect sense if 
\begin{equation}
 M_t^2 \ll M_{Pl}^2, \quad M_s^2 \ll M_{Pl}^2
  \label{eqn:condition-RGflow-original}
\end{equation}
in the UV with $\lambda\to\infty$. The same conclusion holds for all
other terms in the action (\ref{initact}) since all terms which are not 
included in $I_{kin}$ are independent of $\lambda$ and are either marginal
or relevant. The condition (\ref{eqn:condition-RGflow-original}) is
equivalent to (\ref{eqn:condition-RGflow}) and should be considered as a
nontrivial condition on properties of the RG flow in the vicinity of 
$\lambda=+\infty$ in the UV.

\section{Summary and Discussion}
\label{sec:summary}

In this paper, we have studied the dynamics of the projectable
Ho\v{r}ava-Lifshitz (HL) gravity with the $z=3$ scaling in the
ultraviolet (UV), focusing on the limit $\lambda\to\infty$. This limit
for the parameter $\lambda$ is a natural candidate for the UV fixed
point of the renormalization group (RG) flow, if one forbids a ghost
degree of freedom (appearing in the regime $1/3<\lambda<1$) and hopes
that general relativity (GR) (having $\lambda=1$) be recovered at low
energy. Contrary to naive expectations, the system is well behaved in
the limit $\lambda\to\infty$. Indeed, the dynamics can be even simpler
due to the $1/\lambda$ suppression of the coupling between the gravity
and matter sectors. We have analyzed tensor and scalar gravitons in the
FRW universe driven by ``dark matter as integration constant'', and
shown that the amplitudes of quantum fluctuations remain finite. The
theory in the UV with $\lambda\to\infty$ is weakly coupled, provided
that the condition (\ref{eqn:condition-RGflow}) is satisfied.

While we have argued that the theory behaves well in the UV with the
$\lambda\to\infty$ limit, cosmological implication of the result has not
been explored yet. This is because of the lack of our understanding of the
low energy dynamics with the $\lambda\to 1+0$ limit. This limit is the
candidate for an infrared fixed point of the RG flow since GR has the
value $\lambda=1$.

It is known that in the limit $\lambda\to 1+0$, the scalar graviton gets
strongly coupled. Strong coupling itself does not imply loss of
predictability since all coefficients of infinite number of terms in the
perturbative expansion can be written in terms of finite number of
parameters in the action if the theory is renormalizable. However, the
strong coupling implies breakdown of the perturbative expansion in the
scalar graviton sector and, thus, we need nonperturbative analysis. For
spherically-symmetric, static, vacuum configurations, it was shown by
nonperturbative analysis that the limit $\lambda\to 1+0$ is indeed
continuous and recovers GR~\cite{Mukohyama:2010xz}. This result may be 
considered as an analogue of Vainshtein effect and suggests the
possibility that the scalar graviton may safely be decoupled from the
rest of the world, i.e. the tensor graviton and the matter sector, in
the limit $\lambda\to 1+0$. A similar work for super-horizon nonlinear
cosmological perturbations in universes driven by ``dark matter as
integration constant'' is in progress~\cite{Mukohyama:soon}. 
Nonetheless, it is fair to say that our understanding of the fate of the
scalar graviton in the limit $\lambda\to 1+0$ is far from complete.

For this reason, we have not conducted a full analysis of
cosmological implication (e.g. on the CMB spectrum) of the result of
this paper.

Fortunately, the simple scenario in \cite{Mukohyama:2009gg} does not
suffer from the lack of our understanding of the $\lambda\to 1+0$
limit. For example, one can reliably calculate non-Gaussianities in
cosmological perturbations~\cite{Izumi:2010yn}. A scalar field
responsible for (almost) scale-invariant cosmological perturbations acts
as a curvaton: it is sub-dominant at the time of sound horizon exit,
later becomes dominant and finally reheats the universe. The only
property of HL gravity needed for this mechanism is the anisotropic
scaling with $z=3$. (Thus, this mechanism should work also in other
versions of HL gravity~\cite{Blas:2009qj,Horava:2010zj}.) When energy
density of this scalar field is sub-dominant in the early epoch, it is
expected that the only important effect of gravity to the dynamics of
the scalar field is to provide an expanding background. Therefore, if
$\lambda$ runs towards $1$ and GR is recovered during the epoch when the
scalar field is sub-dominant, then the prediction of this scenario does
not depend on details of the behavior of the scalar graviton in the
limit $\lambda\to 1+0$.

An open issue regarding the scenario in \cite{Mukohyama:2009gg} is to
find a mechanism for Lorentz invariance restoration in the matter sector
at low energies. Actually, this issue is shared by HL gravity itself:
even if one omits Lorentz violating terms in the matter sector, these
terms will be generated by radiative corrections from graviton
loops. These terms may be under control provided that $M\ll M_{\rm Pl}$,
where $M$ is the scale at which the anisotropic scaling becomes
important~\cite{Pospelov:2010mp}. Another approach to this problem is to 
enforce a universal Lorentz breaking at all sectors at high energies, while
supersymmetrizing the standard model ensures the restoration of the
Lorentz symmetry at low energies 
\cite{GrootNibbelink:2004za,Xue:2010ih}.

\acknowledgments
Part of this work was done during YITP molecule-type workshop (T-10-05): 
Cosmological Perturbation and Cosmic Microwave Background. The authors 
thank YITP for stimulating atmosphere and warm hospitality. This work was supported by the World Premier International Research Center Initiative (WPI Initiative), MEXT, Japan. S.M. acknowledges the support by Grant-in-Aid for Scientific Research 17740134, 19GS0219, 21111006, 21540278, by Japan-Russia Research Cooperative Program.

\appendix

\section{Example with scalar field}
\label{app:scalar}

In Sec.~\ref{sec:FRW} we have seen that coupling between gravity and
matter sectors is suppressed by $1/\lambda$ and that these sectors
decouple in the limit $\lambda\to\infty$. Motivated by this, in 
Sec.~\ref{sec:perturbation} we have studied a pure gravity system and 
analyzed the evolution of perturbations around the FRW background driven
by the ``dark matter as integration constant''.  In this appendix, in
order to justify this treatment, we consider a simple system of a scalar
field in HL gravity and show that gravity and matter are indeed
decoupled in the limit $\lambda\to\infty$ for linear perturbations.

We consider a single Lifshitz scalar field with the dynamical critical
exponent $z=3$, in accordance with the gravity sector. The dynamics of
the field is described by the action 
\begin{equation}
I_m = \frac{1}{2} \,\int dt \,d^3x \,N \,\sqrt{g}\,\left[\frac{1}{N^2}\,\left(\partial_t\,\varphi-N^i\partial_i\varphi\right)^2+\varphi\,{\cal O}_\phi\,\varphi-2\,V\right]\,,
\label{fieldac}
\end{equation}
where the operator containing the gradients is defined as
\begin{equation}
{\cal O}_\phi \equiv \frac{1}{M_\phi^4}\,\left(D_iD^i\right)^3 - \frac{\kappa_\phi}{M_\phi^2}\,\left(D_iD^i\right)^2 + c_\phi^2\,D_iD^i\,.
\end{equation}
After decomposing the field into zero mode and perturbations as $\varphi
= \phi+\delta \phi$, we vary the background action with respect to the
scale factor and the field, to obtain the equations of motion 
\begin{eqnarray}
&& -\frac{3\,\lambda-1}{2}\,\left(2\,\dot{H}+3\,H^2\right) = \frac{1}{M_{\rm Pl}^2}\,\left(\frac{\dot{\phi}^2}{2}-V\right)\,,\nonumber\\
&&\ddot{\phi}+3\,H\,\dot{\phi}+V'=0\,,
\end{eqnarray}
where the second (Klein-Gordon) equation is a special case of the
energy-nonconservation equation (\ref{encon}), with $Q(t)=0$. This extra
information comes from specifying a field source for the perfect fluid
description. 

For perturbations, we calculate the quadratic action and expand modes in
Fourier space, through 
\begin{equation}
\delta (t, {\vec x}) = \frac{1}{\left(2\,\pi\right)^{3/2}}\,\int d^3k \, {\rm e}^{i\,{\vec k}\cdot {\vec x}}  \,\hat{\delta}(t,{\vec k})\,,
\end{equation}
where $\delta$ represents any scalar degree. The momentum constraint
gives a relation for the nondynamical degree of freedom $B$, 
\begin{equation}
\hat{B} = -\frac{a^2}{k^2}\left[
\frac{3\,\lambda - 1}{\lambda -1}\,\dot{\hat{\zeta}} + \frac{\dot{\phi}}{(\lambda-1)M_{\rm Pl}^2}\,\delta\hat{\phi}
\right]\,.
\end{equation}
After eliminating $B$ by using this relation, the resulting action turns
out to be a coupled system involving $\delta \phi$ and $\zeta$. In order
to analyze the system, we perform the following field redefinition 
\begin{equation}
\psi \equiv a^{3/2} \left(
\begin{array}{c}
\delta \phi\\
\sqrt{\frac{2\,(3\,\lambda-1)}{\lambda-1}}\,M_{\rm Pl}\,\zeta
\end{array}
\right)\,,
\end{equation}
to obtain the canonically normalized action
\begin{equation}
I = \frac{1}{2} \int dt \,d^3k\,\left(\dot{\hat{\psi}}^\dagger \dot{\hat{\psi}} + \dot{\hat{\psi}}^\dagger X \hat{\psi} - \hat{\psi}^\dagger X \dot{\hat{\psi}} - \hat{\psi}^\dagger \Omega^2 \hat{\psi}\right)\,,
\label{actionform}
\end{equation}
where $X = -X^T$ and $\Omega^2 = (\Omega^2)^T$ are both real matrices,
with elements 
\begin{equation}
 X_{11} = X_{22} = 0, \quad
X_{12} = -\frac{\dot{\phi}}
{\sqrt{2\,\left(3\,\lambda-1\right)\left(\lambda-1\right)}\,M_{\rm Pl}}\,,
\end{equation}
\begin{eqnarray}
\left(\Omega^2\right)_{11} &=& \frac{k^6}{M_\phi^4 a^6} + \frac{\kappa_\phi k^4}{M_\phi^2 a^4} + \frac{c_\phi^2 k^2}{a^2} - \frac{3\,V}{2\,M_{\rm Pl}^2\,(3\,\lambda-1)} -\frac{(9\lambda-1)}{4\,M_{\rm Pl}^2 (\lambda-1)(3\lambda-1)}\dot{\phi}^2 +V''\,,\nonumber\\
\left(\Omega^2\right)_{22} &=& \frac{\lambda-1}{3\,\lambda-1}\,\left(\frac{k^6}{M_s^4 a^6} + \frac{\kappa_s k^4}{M_s^2 a^4} - \frac{k^2}{a^2}\right) + \frac{3}{2\,M_{\rm Pl}^2\,(3\,\lambda-1)}\,\left(\frac{\dot{\phi}^2}{2}-V\right)\,,\nonumber\\
 \left(\Omega^2\right)_{12} &=& -\frac{V'}{\sqrt{2\,\left(3\,\lambda-1\right)\left(\lambda-1\right)}\,M_{\rm Pl}}\,.
\end{eqnarray}
(For a general formalism to quantize coupled bosons, see e.g. 
\cite{Nilles:2001fg}.) For the action (\ref{actionform}), the couplings
between the two degrees of freedom are suppressed by $1/\lambda$, and an
initial adiabatic vacuum state can be defined unambiguously at early times 
\begin{equation}
\psi_1= \frac{M_\phi\,a^{3/2}}{\sqrt{2}\,k^{3/2}}\,{\rm e}^{-i\frac{k^3}{M_\phi^2}\int\frac{dt}{a^3}} \,,\qquad
\psi_2 = \frac{3^{1/4}\,M_s}{\sqrt{2}\,k^{3/2}}\,{\rm e}^{-i\frac{k^3}{\sqrt{3}\,M_s^2}\int\frac{dt}{a^3}}\,.
\end{equation}
In other words, at leading order in $1/\lambda$ expansion, the gravity
sector ($\zeta$) is once again, decoupled from the matter sector
($\delta\phi$). In the UV regime with $\lambda\to\infty$, the solutions
for both physical mode functions have constant amplitudes 
\begin{equation}
\delta\phi = \frac{M_\phi}{\sqrt{2}\,k^{3/2}}\,{\rm e}^{-i\frac{k^3}{M_\phi^2}\int\frac{dt}{a^3}} \,,\qquad
\zeta = \frac{M_s}{2\times 3^{1/4}\,k^{3/2}\,M_{\rm Pl}}\,{\rm e}^{-i\frac{k^3}{\sqrt{3}\,M_s^2}\int\frac{dt}{a^3}}\,.
\end{equation}

\end{document}